\documentclass[]{iac}
\usepackage{xcolor}
\usepackage{breqn}
\usepackage{booktabs}
\usepackage[nolist, nohyperlinks]{acronym}
\usepackage[hyphens]{url}
\usepackage{url}
\usepackage[hidelinks]{hyperref} 
\hypersetup{breaklinks=true}
\usepackage{times}  
\usepackage{helvet}  
\usepackage{courier}  
\usepackage[hyphens]{url}  
\usepackage{graphicx} 
\usepackage{cite}
\usepackage{caption} 
\DeclareCaptionStyle{ruled}{labelfont=normalfont,labelsep=colon,strut=off} 
\usepackage{algorithm}
\usepackage{algorithmic}

\usepackage{times}
\usepackage{epsfig}
\usepackage{graphicx}
\usepackage{amsmath}
\usepackage{amssymb}
\usepackage{array}
\usepackage[hidelinks]{hyperref}

\usepackage{longtable,tabularx}
\usepackage{amsmath}
\usepackage{graphicx}
\graphicspath{ {./images/} }
\usepackage{caption}
\usepackage{subcaption}

\makeatletter
\def\tagform@#1{\maketag@@@{(\ignorespaces#1\unskip\@@italiccorr)}}
\makeatother

\usepackage[acronym,nonumberlist]{glossaries}
\makeglossaries
\newacronym{MOCAT-3}{MOCAT-3}{MIT Orbital Capacity Assessment Tool - 3 species}
\glsaddall

\usepackage{hyperref}
\usepackage{newfloat}
\usepackage{listings}
\newcolumntype{L}[1]{>{\raggedright\arraybackslash}m{#1}}
\newcolumntype{C}[1]{>{\centering\arraybackslash}m{#1}}
\newcolumntype{R}[1]{>{\raggedleft\arraybackslash}m{#1}}
\begin{document}

\IACpaperyear{22}
\IACpapernumber{A6.IPB}  
\IACconference{73}
\IAClocation{Paris, France}
\IACdate{18-22 September 2022}
\IACcopyrightA{2022}{the authors}

\title{A Dynamical Systems Analysis of the Effects of the Launch Rate Distribution on the Stability of a Source-Sink Orbital Debris Model
}

\IACauthor{Celina Pasiecznik}{1}{}
\IACauthor{Andrea D’Ambrosio}{2}{}
\IACauthor{Daniel Jang}{3}{}
\IACauthor{Richard Linares}{4}{}

\IACaffiliation{SM Candidate, Department of Aeronautics and Astronautics, Massachusetts Institute of Technology, MA 02139, USA.}{cpasiecz@mit.edu}{1}
\IACaffiliation{Postdoctoral Associate, Department of Aeronautics and Astronautics, Massachusetts Institute of Technology, MA 02139, USA.}{andreada@mit.edu}{2}
\IACaffiliation{PhD Candidate, Department of Aeronautics and Astronautics, Massachusetts Institute of Technology, MA 02139, USA.}{djang@mit.edu}{1}
\IACaffiliation{Rockwell International Career Development Professor, Associate Professor of Aeronautics and Astronautics, Department of Aeronautics and Astronautics, Massachusetts Institute of Technology, MA 02139, USA.}{linaresr@mit.edu}{4}

\abstract{Future launches are projected to significantly increase both the number of active satellites and aggregate collision risk in Low Earth Orbit (LEO). Ensuring the long-term sustainability of the space environment demands an accurate model to understand and predict the effect of launch rate distribution as a major driver of the evolution of the LEO orbital population. In this paper, a dynamical systems theory approach is used to analyze the effect of launch rate distribution on the stability of the LEO environment. A multi-shell, three-species source-sink model of the LEO environment referred to as MOCAT-3 for MIT Orbital Capacity Assessment Tool - 3 Species, is used to study the evolution of the species populations.
The three species included in the model are active satellites, derelict satellites, and debris. Each shell is modeled by a system of three equations, representing each species, that are coupled  through coefficients related to atmospheric drag, collision rate, mean satellite lifetime, post-mission disposal probability, and active debris removal rate. The major sink in the model is atmospheric drag, whereas the only source apart from collision fragments is the launch rate, making it the critical manageable factor impacting the orbital capacity.
Numerical solutions of the system of differential equations are computed, and an analysis of the stability of the equilibrium points is conducted for numerous launch rate distributions. The stability of the equilibrium points is used to test the sensitivity of the environment to run-away debris growth, known as Kessler syndrome, that occurs at the instability threshold. Various bounding cases are studied from business-as-usual launch rates based on historic launch data, to high launch rates wherein a fraction of the satellite proposals filed with the International Telecommunication Union (ITU) are launched. An analysis of the environment's response to perturbations in launch rate and debris population is conducted. The maximum perturbation in the debris population from the equilibrium state, for which the system remains in a stable configuration, is calculated. Plots of the phase space about the equilibrium points are generated. The results will help to better understand the orbital capacity of LEO and the stability of the space environment, as well as provide improved guidelines on future launch plans to avoid detrimental congestion of LEO.
}{Source-Sink, Launch Rate, System Dynamics, Kessler Syndrome, Debris Evolutionary Model}

\maketitle

\section{Introduction}
\subsection{Motivation}

\begin{figure} 
    \centering
    \includegraphics[width=0.5\textwidth]{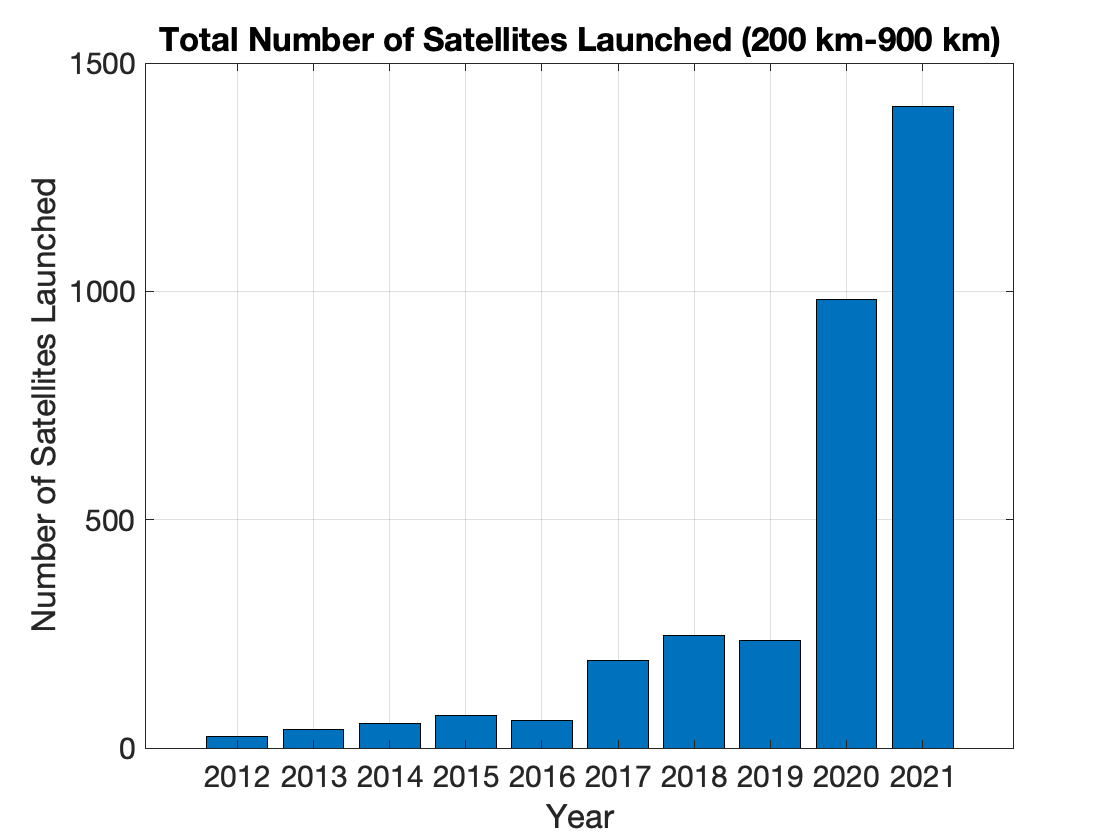}
    \caption{Number of objects launched to $200-900$ km altitudes over the past decade \cite{UCS}.}
    \label{lastten}
\end{figure}

The unprecedented launch rate of satellites into LEO may have severe consequences on the stability of the orbital environment for decades to come.
In the $200-900$ km altitude range alone, the number of satellites launched per year has increased over the last decade as shown in Figure \ref{lastten}. 
Companies such as Amazon and SpaceX, have announced plans to launch constellations of thousands of satellites into LEO over the next decade \cite{davenport_2020} with hundreds of satellites already launched. 
A rising launch rate will increase orbital congestion which in turn raises the chance of debris-generating collisions. A growing debris population can have catastrophic consequences on space missions.  
It is important to study how increased launch activities affect the evolution of the orbital environment and the production of debris, as well as how increased debris populations affect the stability of LEO.

\subsection{Literature Review} 
Various analytic models of the orbital environment have been proposed in the literature that make use of differential equations to represent the evolution of the number of objects in space and their interactions. 
Some of these models have also been used to test the LEO environment's sensitivity to run-away debris growth. 
This run-away debris growth is known as Kessler syndrome wherein the congestion of the orbital environment is large enough to cause a chain reaction of debris generation. 
To study such debris growth caused by collisions, Kessler and Cour-Palais developed a source-sink model to predict detrimental debris population growth \cite{kessler}.
Furthermore, Talent \cite{Talent} used one ODE to represent the total number of objects in space and studied various evolutionary cases for different launch rates to look for catastrophic behaviour. 
Zhang et al. \cite{Zhang} developed a model using partial differential equations and solved the equations numerically to study the long-term evolution of the space debris environment. 
A dynamical systems analysis was conducted by Drmola and Hubik \cite{Drmola} in which three different classes of debris were used to study various debris accumulation scenarios and whether they lead to Kessler syndrome. 

The MOCAT model was inspired the three population model JASON given in reference \cite{Lewis2009TheModel}. The MOCAT-3 model has been used to calculate the intrinsic capacity of LEO in reference \cite{capacity}. The model was extended to included a differentiation between slotted and unslotted satellites in reference \cite{AAS}. Here, a brief overview of the MOCAT-3 model is given, and various numerical analysis are conducted using the model for the study of launch rates and debris generation.

\subsection{Paper objectives}
The objective of this paper is to study the evolution of the LEO environment for various launch rate distributions and to determine the stability of the environment in response to perturbations in the debris population. 
The intent of finding the stability boundary is to determine how large the debris population can grow before run-away debris growth, referred to as Kessler syndrome, occurs.
 
\subsection{Paper structure}
The remainder of the paper is arranged as follows: Section \ref{methods} gives an overview of the source-sink model and methods used for conducting the stability analysis; Section \ref{launches} introduces various launch rate distribution cases; Section \ref{sec:results} presents the equilibrium solutions and the results of perturbations to these solutions; Section \ref{kessler} shows the analysis of the instability threshold where Kessler syndrome occurs; and finally Section \ref{sec:conclusions} remarks on the meaning of the analysis and draws conclusions about the future of the orbital environment.

\section{Methods}
\label{methods}
\subsection{MOCAT-3 Model} 
The MOCAT-3 Model was developed in reference \cite{capacity} wherein a detailed description of each parameter is given. 
Here we provide a brief overview of the model and the key equations describing the evolution of each species. 

MOCAT-3 is a probabilistic source-sink model with three species: active satellites (S), derelict satellites (D) and debris (N). The orbital environment within the altitude range of $200-900$ km is divided into $20$ spherical orbital shells with a shell thickness of $35$ km, represented by the variable $d$. 
The evolution of each species is represented by a set of differential equations per shell $\{\dot{S}(h)$, $\dot{D}(h)$, $\dot{N}(h)\}$, where $h$ is a value from $1$ to $20$ indicating the shell number. 
Shell with $h=1$ is the lowest shell for the altitude range $200-235$ km and shell with $h=20$ is the highest shell for the altitude range $865-900$ km. 
We assume each object has a near-circular orbit. 
The launch rate per year is represented by $\lambda(h)$ and only appears as a source in $\dot{S}$. New active satellites appear instantly in their orbital shell $h$ and do not cross through lower shells. 
Dropping the explicit dependence on shell number, the set of coupled ordinary differential equations representing the evolution of each species is given by equations (\ref{sdot}), (\ref{ddot}), (\ref{ndot}).
\begin{dmath}
\label{sdot}
\dot{S}=\lambda-S / \Delta t -\phi_{SN}(\delta+\alpha)N S -\phi_{SD}(\delta+\alpha)D S-\alpha_{a} \phi_{SS} S^{2} \quad
\end{dmath}
\begin{dmath}
\label{ddot}
\dot{D}=\frac{(1-P) S}{\Delta t}+\phi_{SD} \delta D S+\phi_{SN} \delta N S-\phi_{DN} N D-\phi_{DD}D^2 +\frac{D_{+} v_{D+}}{d}-\frac{D v_{D}}{d}
\end{dmath}
\begin{dmath}
\label{ndot}
\dot{N}= K0_{SN} \phi_{SN} \alpha N S
+ K0_{SD} \phi_{SD} \alpha D S
+ K0_{DN} \phi_{DN} N D
+ K0_{DD} \phi_{DD} D^2
+ \alpha_a K0_{SS} \phi_{SS} S^2
+ K0_{NN} \phi_{NN} N^2
+\frac{N_{+} v_{N+}}{d} - \frac{N v_{N}}{d}
\end{dmath}
Here $D_+$ and $N_+$ refer to the populations of $D$ and $N$ in the shell directly above the current shell, namely $D_+=D(h+1)$ and $N_+=N(h+1)$. 
For the highest shells, we assume these parameters represent the current shell: $D_+=D(h)$ and $N_+=N(h)$, for reasons given in section \ref{equilibriumcalc}. 
In general, once an object is in orbit it can only flow into lower altitude shells and not into upper shells. The de-orbiting of objects from higher shells to lower shells is dictated by a static exponential model for the atmospheric density described in \cite{capacity}. Only derelict and debris objects are assumed to de-orbit from atmospheric drag effects as active satellites are assumed to have station-keeping capabilities that counter-act these drag effects.
These drag effects are accounted for in the terms containing the variable $v$ which represent the change in the semi-major axis.
Active satellites $S$ can become derelict $D$ or debris $N$ through collisions but no species can become an active satellite $S$ for which the only source is $\lambda$. 
Furthermore, active satellites directly exit the environment at a rate of $1/\Delta t$ with a success probability of $P$, the rest becoming derelict satellites. 
The lifetime of each active satellite is taken as $\Delta t = 5$, whereas the probability of successful post-mission disposal is $P=0.95$.
The number of fragments created during collisions between the species is determined by the NASA standard break-up model \cite{krisko2011proper} which determines the values of $K0$ and $\phi$. These variables are specified for each type of collision between all species combinations.
For the collision model, the average mass, area, and diameter values used for each species was taken from reference \cite{somma}, and are shown in Table \ref{tablesomma}. 
The variables $\delta$, $\alpha$, $\alpha_a$ set the proportionality of collisions that become debris objects. Specifically, $\delta=10$
gives the ratio of collisions that produce disabling versus lethal debris, $\alpha=0.2$ is the fraction of derelict and debris objects that an active satellite fails to avoid, and $\alpha_a=0.01$ is the fraction of active satellites that another active satellite fails to avoid. 
This section has summarized the parameters of the MOCAT-3 model which was used for the numerical analysis given in the rest of the paper.

\begin{table}
\centering
\begin{tabular}{llll}\toprule
{} & Active & Derelict & Debris \\ \midrule
Mass (kg) & 223 &223 &0.640\\
Area ($m^2$) &1.741 &1.741 &0.020\\
Diameter (m) &1.490 &1.490&0.180\\
\bottomrule
\end{tabular}
\caption{Physical characteristics of each species.}
\label{tablesomma}
\end{table}

\subsection{Equilibrium Solutions and Stability}
\label{equilibriumcalc}
The equilibrium points for the set of coupled differential equations (\ref{sdot}), (\ref{ddot}), (\ref{ndot}) were solved for each shell by finding the population of each type of species $\{S_{eq}(h),D_{eq}(h),N_{eq}(h)\}$ for which the differential equations equal zero:
$$
\dot{S}=0, \hspace{4pt} \dot{D}=0, \hspace{4pt}\dot{N}=0.
$$
For this set of values $\{S_{eq}(h),D_{eq}(h),N_{eq}(h)\}$, the sources and sinks of the environment balance each other and the system is in equilibrium. 
A change in the launch rate generates a new set of equilibrium solutions because the launch rate is a major source in the active satellites population.
Since each differential equation (\ref{sdot}), (\ref{ddot}), (\ref{ndot}) has degree $2$, the set of $3$ coupled equations has $2^3=8$ equilibrium solutions per shell. We eliminate solutions that are purely imaginary or that contain a real negative part, as these are considered non-physical solutions for the species populations. For the launch rate cases we studied, we found that each shell has two sets of positive, real-valued equilibrium solutions $\{S_1(h),D_1(h),N_1(h)\}$ and $\{S_2(h),D_2(h),N_2(h)\}$, with one solution set having a larger number of active satellites than the other: $S_1 > S_2$. 
We used this solution set $\{S_1(h),D_1(h),N_1(h)\}$ as the influx populations to the next lower shell $\{S_1(h-1),D_1(h-1),N_1(h-1)\}$.
For the highest shell, we assumed the influx of objects from higher altitudes was equaled to the outflow of objects from that shell. This assumption may differ from reality as there are many objects located above $900$ km but it is difficult to measure how many of these objects would de-orbit due to atmospheric drag at such high altitudes per year.
By finding the equilibrium solutions $\{S,D,N\}$ per shell starting with the highest altitude shell and ending with the lowest altitude shell, we guarantee the equilibrium of the entire orbital environment in the 200km-900km altitude range because each shell depends only on the species population within that shell and directly above it. 
Our model does not contain any flows from lower shells into higher shells so the lower shells do not affect the equilibrium of the higher shells.

\begin{figure} 
    \centering
    \includegraphics[width=0.5\textwidth]{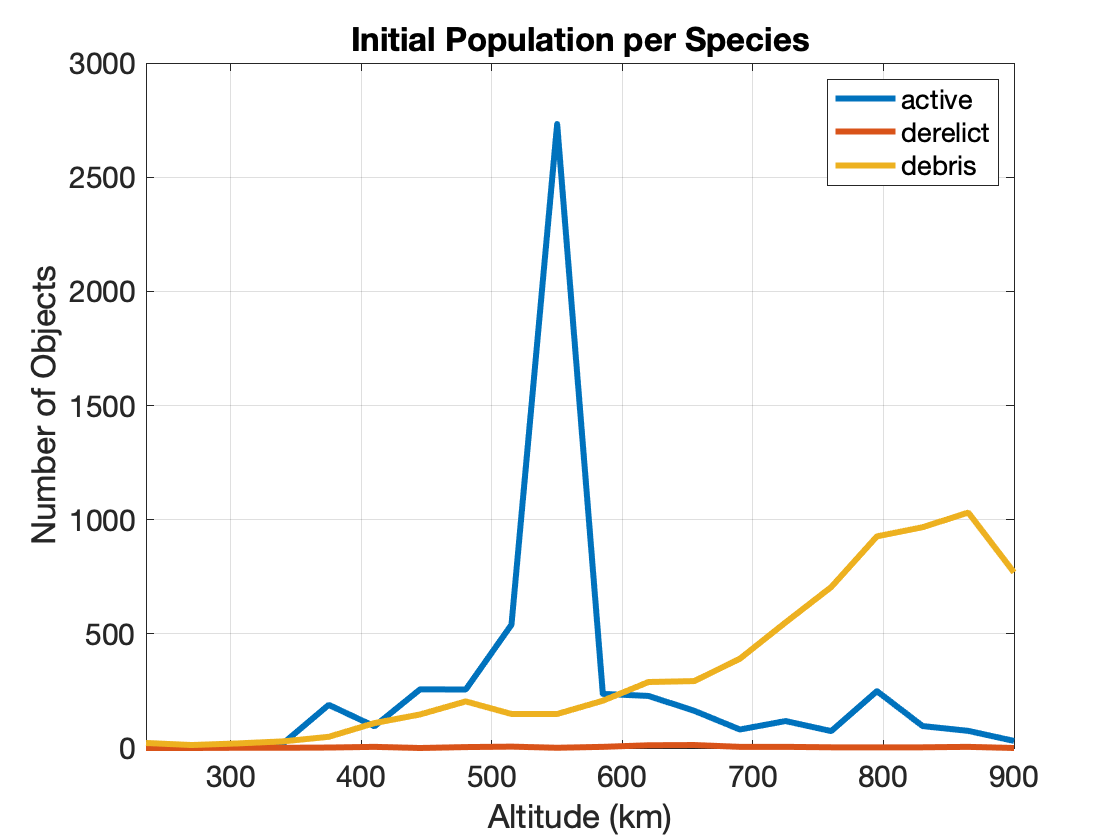}
    \caption{Population of each species from TLE data as of August 2022.}
    \label{initialfig}
\end{figure}

For a given launch rate and set of initial conditions $\{S_{i},D_{i},N_{i}\}$, we integrate the differential equations (\ref{sdot}), (\ref{ddot}), (\ref{ndot}) with respect to time to find the number of years required for the source-sink model to reach equilibrium.
The initial conditions we used was based on Two Line Element data from space-track.org \footnote{The TLE catalog was downloaded from space-track.org, accessed on August 18th, 2022.}.
We used the process of reference \cite{AAS} that classified each object as an active satellite, derelict satellite, or as debris according to its mass, diameter and area. We note that unlike reference \cite{AAS}, we do not distinguish between slotted and unslotted satellites, thus we combined these two populations for the active satellite population. The initial populations of each species are displayed per altitude shell in Figure \ref{initialfig}.

\subsection{Basin of Attraction}
\emph{Basin of Attraction Definition:} ``The set of points in the space of system variables such that initial conditions chosen in this set dynamically evolve to a particular attractor" \cite{mathworld}. 

In our model, an attractor is a stable equilibrium point.
Once the source-sink model is at a steady equilibrium state, we studied how the model reacts to perturbations in $\{S,D,N\}$ from equilibrium. In particular, we studied how a sudden increase in debris $N$ alters the orbital environment by looking at how the system evolved after the perturbation and whether or not it tended back toward equilibrium.
We found the basin of attraction about the stable equilibrium point for various initial conditions of debris $N$. 
One of the reasons we decided to focus on analyzing the stability of the orbital environment with respect to changes in the species of debris rather than changes in active or derelict satellites, is the exact amount of debris currently in LEO is unknown. Another reason is we wanted to study the instability threshold for debris creation known as `Kessler syndrome' in which the amount of debris in the orbital environment is numerous enough that it continuous to generate more and more debris, creating a chain reaction. To visualize the basin of attraction, we plotted phase space diagrams.

\section{Launch Rate Distributions}
\label{launches}
\emph{Launch Rate Distribution Definition:} The number of active satellites launched into orbit per year per altitude shell. 

We used various launch rate distributions to study the stability of the LEO environment. The two types of launch rate distributions that we studied are \emph{static} and \emph{dynamic} launch rates. 
A static launch rate represents a constant influx of active satellites per altitude shell per year for a given number of years. 
A dynamic launch rate represents a variable launch rate per altitude shell per year. 
For each launch rate, a unique set of equilibrium solutions is found. Thus, a static launch rate is used to study the behaviour of the system of equations with respect to the equilibrium solutions. However, a varying launch rate represents a more realistic scenario since the number of satellites launched per year has changed drastically over the past few decades. A dynamic launch rate was studied as a separate case wherein the equilibrium solutions changed with variations in the launch rate.

\subsection{Static Launch Rate}
We studied two cases of static launch rates. For the first case we used the maximum number of satellites launched in one year per altitude shell over the past ten years. The second case, we used `As Received' filings database from the International Communications Union (ITU).

\subsubsection{\emph{Case 1: Past Launch Rates}}
\label{case1}

\begin{figure} 
    \centering
    \includegraphics[width=0.5\textwidth]{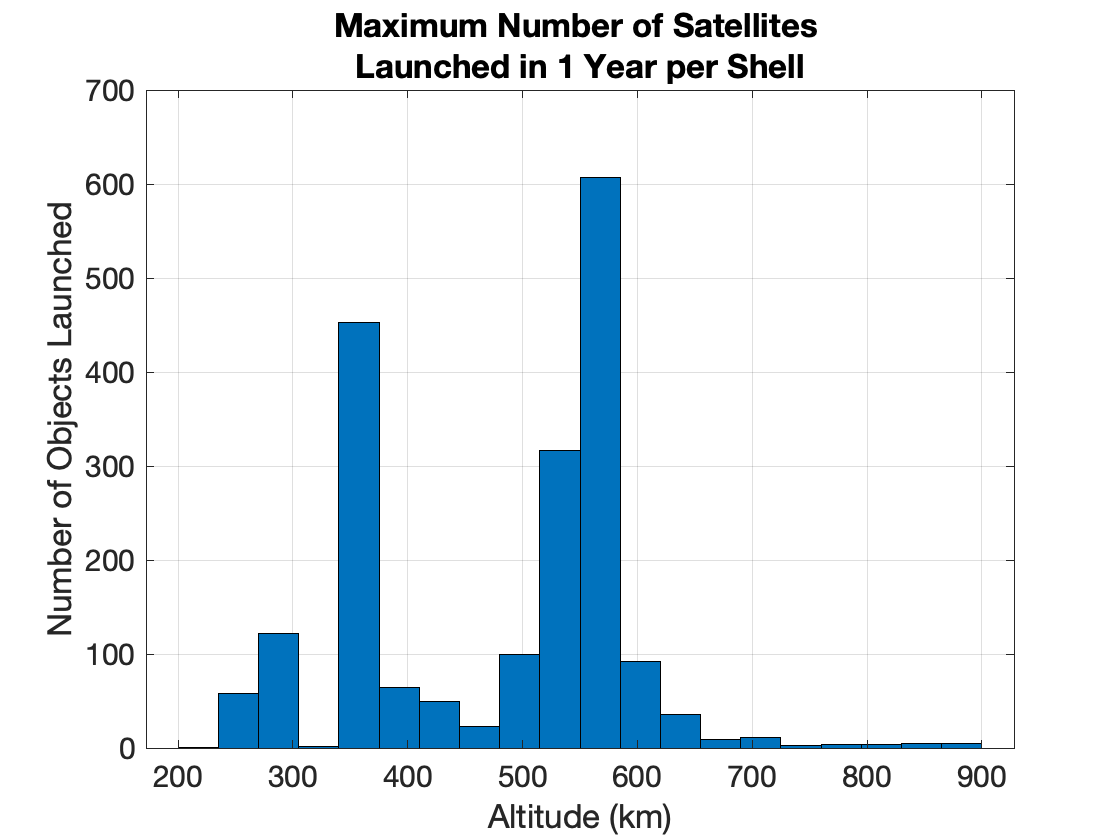}
    \caption{Maximum launch rate over the last ten years per altitude shell \cite{UCS}.}
    \label{unionfig}
\end{figure}

We used the maximum number of satellites launched within one year over the past ten years for each altitude shell. We use the maximum historic launch rate per shell instead of using the launch rate from a specific year because particular years have a low number of launches to certain shells that doesn't represent the general behaviour for launch cadence.
This launch rate distribution allowed us to analyze the stability of the current LEO environment to see if current launch activities are sustainable or if we are already in danger of run-away debris growth. The number of satellites launched into each altitude shell was taken from the Union of Concerned Scientists database \cite{UCS}, and is displayed in Figure \ref{unionfig}. 

\subsubsection{\emph{Case 2: ITU Filings}}
\label{case2}

\begin{figure} 
    \centering
    \includegraphics[width=0.5\textwidth]{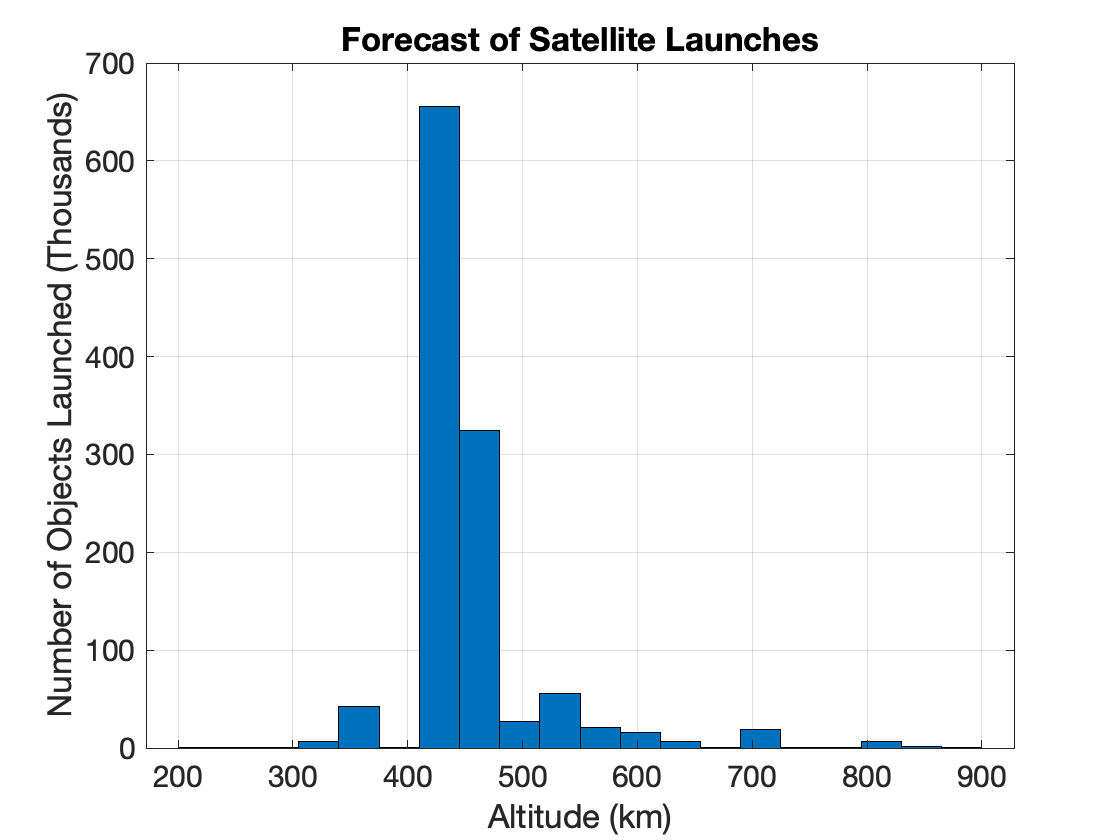}
    \caption{`As Received' ITU filings of satellite notices \cite{itu}.}
    \label{itufig}
\end{figure}

The `As Received' ITU filings database \cite{itu}, is a list of satellite notices filed with the ITU that have not yet been reviewed or published by the ITU. 
It should be noted that the ITU states this database is not regulated. However, this database allows for some forecasting of satellite launches over the next few years. 
Each filing includes the altitude and number of satellites that an organization intends to launch. 
We have filtered through this database to eliminate duplicate filings made by the same organization. We used the average of the apogee and perigee altitude to bin the satellites into altitude shells. The number of satellites forecast to be launched into each shell is shown in Figure \ref{itufig}.
These filings are valid for several years and the deployment of a satellite or constellation of satellites into orbit can take over a year. Thus, in using ITU filings to estimate a launch rate per year, we divided the total number of satellites in each shell by a number of years $n$. 

\subsection{Dynamic Launch Rate}
We studied a dynamic launch rate as a third launch rate distribution. 
As shown in Figure \ref{lastten}, the number of objects launched into orbit each year has not remained constant. 
By using a dynamic launch rate, we can represent such a change in launch rate per year.

\subsubsection{Case 3: Varying Launch Rate per Year}
\label{case3}
\begin{figure} 
    \centering
    \includegraphics[width=0.5\textwidth]{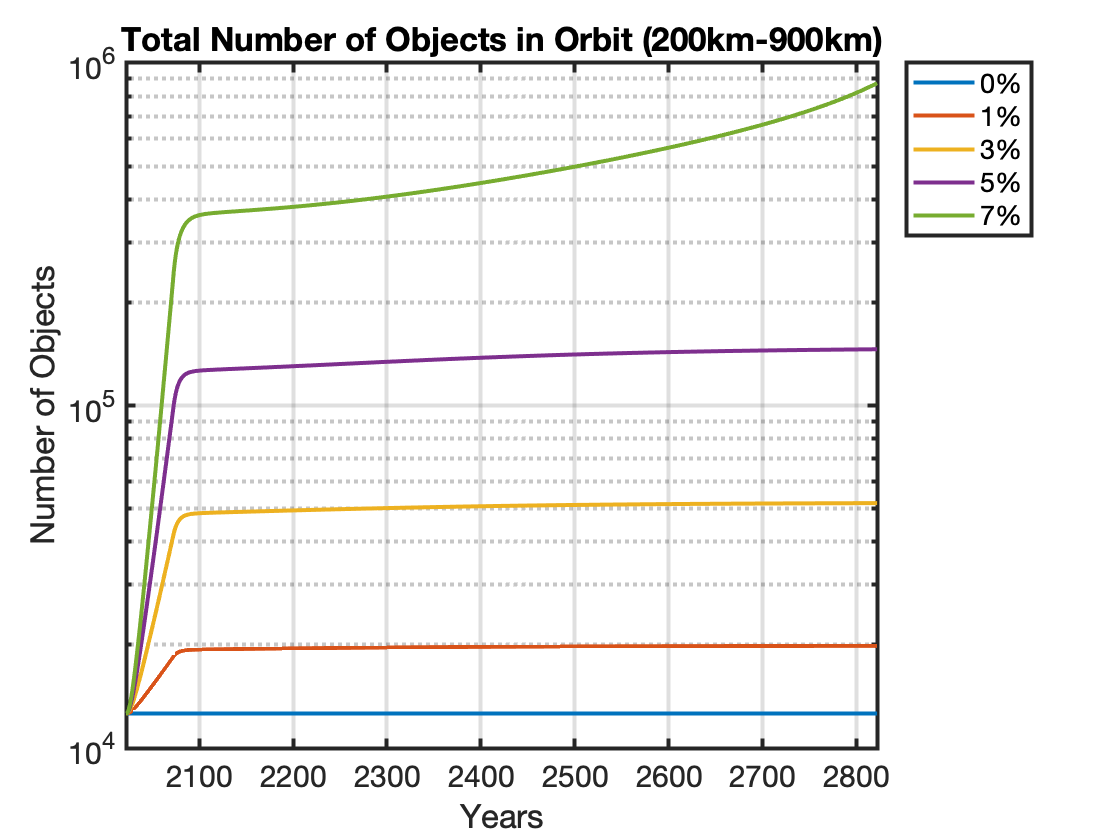}
    \caption{Total number of objects in orbit for launch rates growing by 0-7\% per year for 50 years and then remaining constant for 800 years.}
    \label{dynamic}
\end{figure}

We followed a similar approach to \cite{Talent} in modeling a dynamic launch rate. We took the launch rate displayed in Figure \ref{unionfig} as the base case and then increased this launch rate by 0\%,1\%,3\%,5\%, and 7\% each year for $50$ years. Then we set the launch rate to be constant at the rate calculated at the end of the $50$ years. We let the environment evolve for another $800$ years at this constant launch rate. The total number of objects per year for each incremental launch rate is displayed in Figure \ref{dynamic}. The total number of launched satellites and species populations at the end of the 800 years for each incremental launch rate is given in Table \ref{tabledynamic}. As can be seen in the table, the various  percentage increases in launch rate for the first 50 years produces drastically different total number of objects in the environment at the end of the simulation. For example, a 7\% increase in launch rate produces a total number of objects at the end of the 800 year simulation that is about two orders of magnitude larger than the total produced by a 1\% increase in launch rate. This numerical analysis shows how a few percent difference in increasing launch rate per year creates large differences in the population of each species when propagated over time.

\begin{table}
\centering
\begin{tabular}{|p{0.7cm}||p{0.9cm}|p{0.9cm}|p{1cm}|p{0.9cm}|p{1.1cm}|} \toprule
\% Increase & Launch $\lambda$& Active S & Derelict D& Debris N & Total \\ \midrule
0\% & 1965 &9820&	1383&	1402&	12606 \\
1\% & 3232&	16146&	1984&	1743&	19873\\ 
3\% & 8614&	42976&	4621&	4317&	51914 \\
5\% & 22533	&111830	&12387&	22061&	146270\\
7\% & 57883	&270980&	60571&	741220&	1072800 \\
\bottomrule
\end{tabular}
\caption{Population of each species and total number of objects in orbit at the end of a period of constant growth rate in launch as shown in Figure \ref{dynamic}.}
\label{tabledynamic}
\end{table}

\section{Results and Discussion}
\label{sec:results}
\subsection{Equilibrium Solutions}
\begin{figure} 
    \centering
    \includegraphics[width=0.5\textwidth]{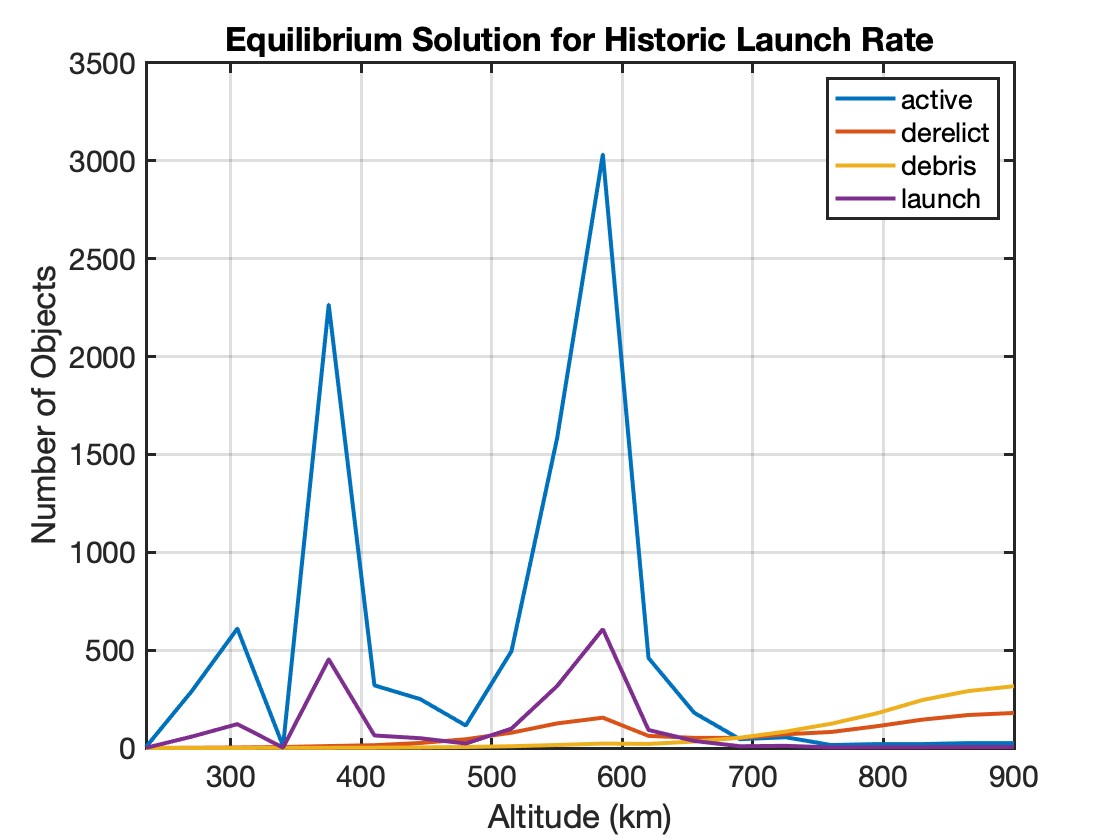}
    \caption{Equilibrium solutions per species for the constant launch rate case \ref{case1}.}
    \label{case1equil}
\end{figure}

For each launch rate, we determined the set of equilibrium solutions.
We computed the equilibrium solutions per shell for the `business as usual' case \ref{case1}; the results for which are presented in Figure \ref{case1equil}. Since equilibrium solutions exist, we can conclude that our current launch activity is sustainable for the $200-900$ km altitude range and that run-away debris growth will not occur if launch rates remain at these levels.

\begin{figure} 
    \centering
    \includegraphics[width=0.5\textwidth]{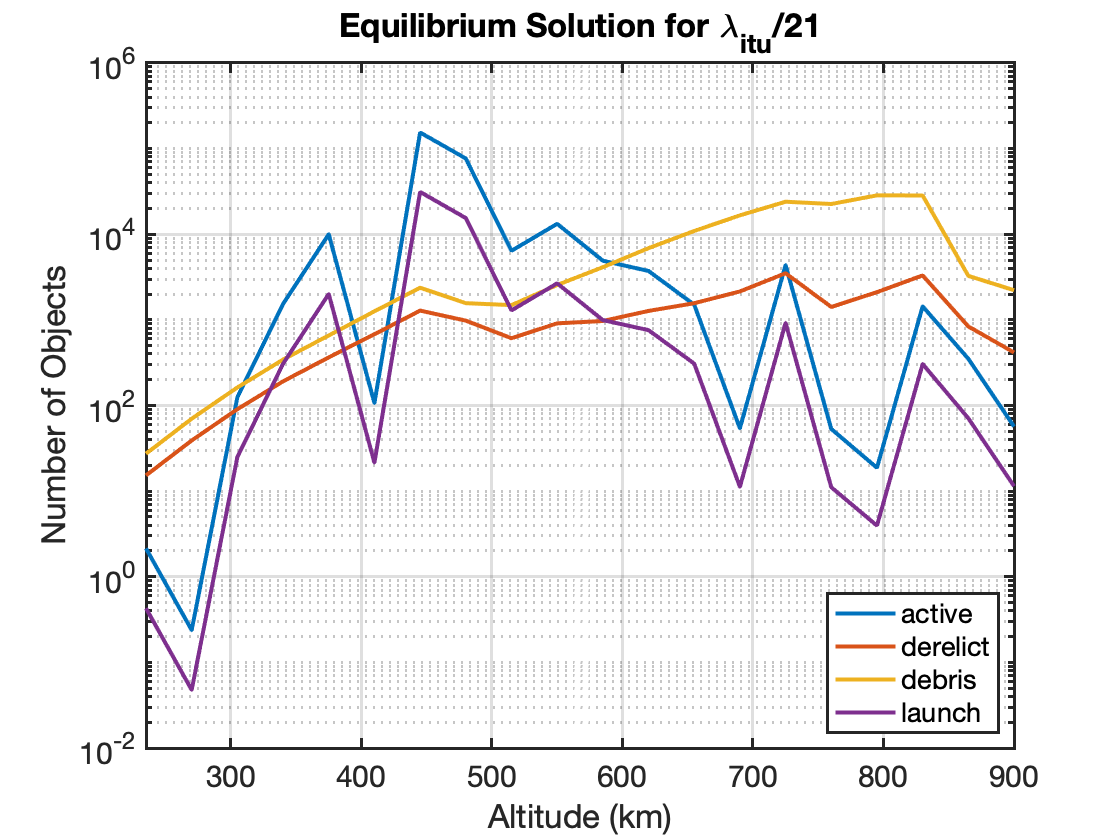}
    \caption{Equilibrium solutions per species for a constant launch rate proportional to ITU filings given in case \ref{case2}.}
    \label{ituequil}
\end{figure}

We also computed the equilibrium solutions for case \ref{case2} where $\lambda_{itu} /n$ with $n=7$ years and found that no positive, real-valued solutions existed.
This may seem alarming but realistically we should not assume that all the satellite notices filed will be used. Only a fraction of the satellite notice filings will actually be launched. We increased $n$ to $n=21$, and found a set of equilibrium points for each shell as shown in Figure \ref{ituequil}.

For the dynamic launch case \ref{case3}, positive, real-valued equilibrium solutions could only be found for the 1\% increase in launch rate per year. A summary of total populations of each species at equilibrium for each launch case is given in Table \ref{tableequil}.  

\begin{table*}
\centering
\begin{tabular}{llllll}\toprule
Rate & Total Launch $\lambda$& Total Active S & Total Derelict D& Total Debris N & Total All \\ \midrule
Case 1 & 1965 &9820&	1383&	1402&	12606 \\
Case 2 & 56410	&278847&	22704&	157870&	459421\\ 
Case 3: 1\% & 3232&	16145&	2292&	4281&	22719 \\
\bottomrule
\end{tabular}
\caption{Total population of each species at equilibrium for various launch rate distributions.}
\label{tableequil}
\end{table*}

Overall, the amount of active satellites $S$ at equilibrium is proportional to the number of satellites launched. The amount of debris $N$ and derelict satellites $D$, however, vary with respect to the amount of active satellites, with higher shells acquiring a larger number of derelict and debris objects because the sink caused by atmospheric drag removes less objects per year at higher altitudes due to the lower atmospheric density. 


\subsection{Stability Analysis}
The stability of the equilibrium solutions found for Case 1 (\ref{case1}), Case 2 (\ref{case2}) and Case 3 (\ref{case3})  at 1\% increment only, was determined by computing the eigenvalues of the solutions. All eigenvalues were found to be negative indicating all of these equilibrium solutions are stable. As an example, Table \ref{eigenvalues} displays the eigenvalues for Case 2 launch rate of section \ref{case2}. 

\begin{figure} 
    \centering
    \includegraphics[width=0.5\textwidth]{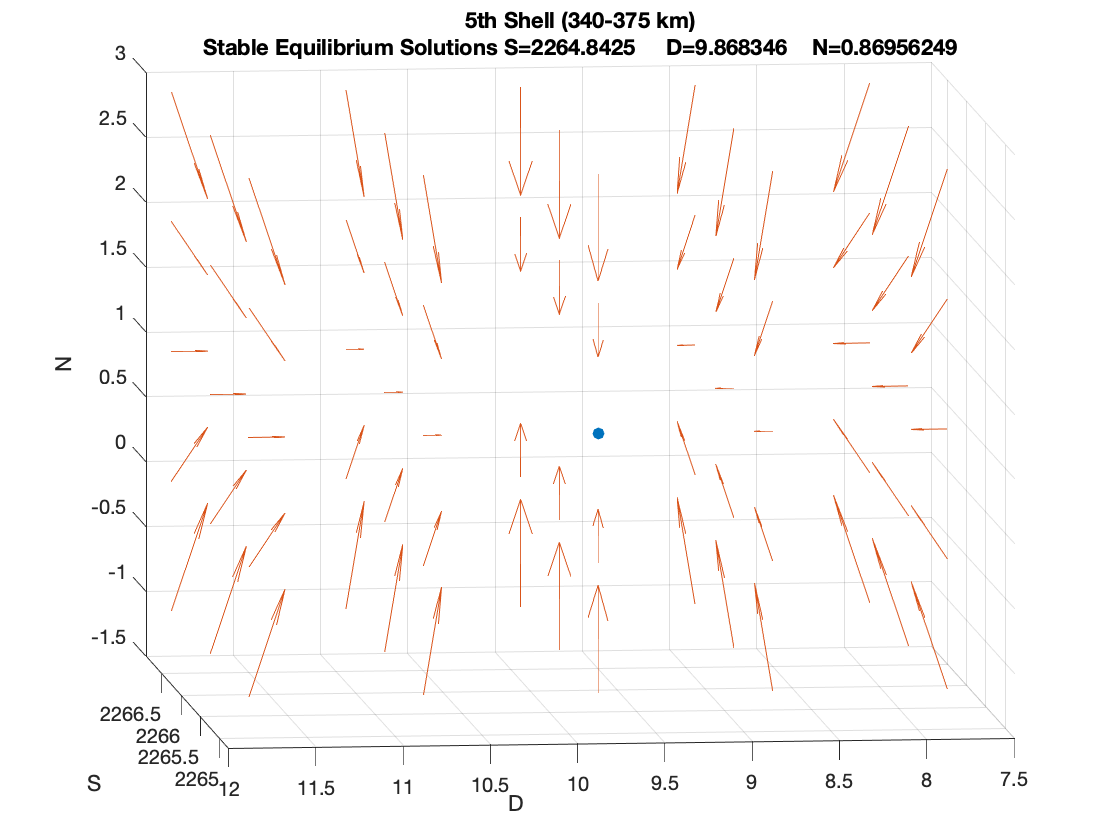}
    \centering
    \includegraphics[width=0.5\textwidth]{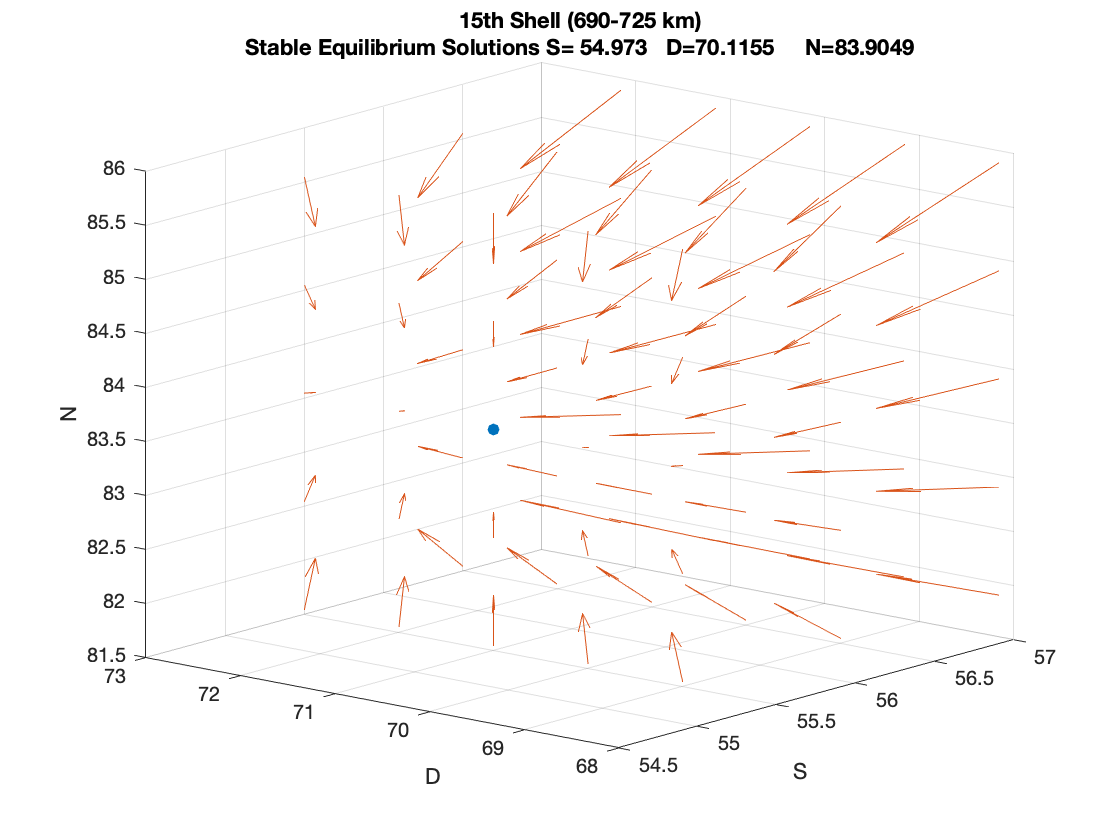}
    \caption{Phase portraits about the stable equilibrium state at different altitudes.}
    \label{phaseport}
\end{figure}

\begin{table}
\centering
\begin{tabular}{lll} \toprule
Active S & Derelict D& Debris N \\ \midrule
-0.200&	-1.028	&-0.212\\
-221.167&	-0.591&	-0.211\\
-885.271&	-0.432&	-0.211\\
-345.598&	-0.350&	-0.200\\
-149.086&	-0.253&	-0.200\\
-86.341	&-0.119&	-0.209\\
-37.246	&-0.089	&-0.208\\
-70.472&	-0.077&	-0.208\\
-36.780	&-0.056&	-0.205\\
-19.147	&-0.043&	-0.200\\
-17.605	&-0.001&	-0.203\\
-10.008	&-0.003&	-0.205\\
-9.180	&-0.004&	-0.201\\
-5.522&	-0.007&	-0.201\\
-4.784&	-0.017&	-0.201\\
-3.129&	-0.024&	-0.204\\
-2.350&	-0.026&	-0.203\\
-1.306&	-0.027&	-0.203\\
-1.776&	-0.027&	-0.201\\
-0.777&	-0.147&	-0.202\\
\bottomrule
\end{tabular}
\caption{The eigenvalues of each population for the equilibrium solutions displayed in Figure \ref{ituequil}.}
\label{eigenvalues}
\end{table}
Additionally, we used phase portraits to depict the phase space about the stable equilibrium point per shell. The phase portraits for the equilibrium points of the Case 1 launch rate, shown in Figure \ref{case1equil}, are displayed for two altitude shells in Figure \ref{phaseport}. 

\begin{figure} 
    \centering
    \includegraphics[width=0.5\textwidth]{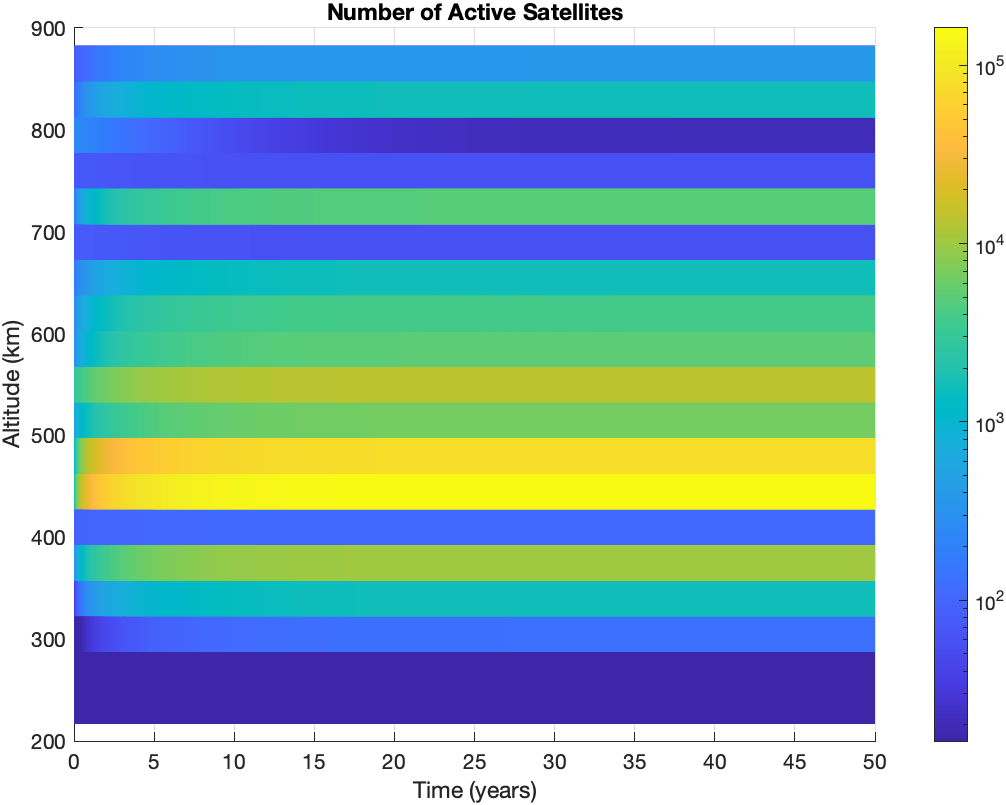}
    \caption{Number of active satellites in orbit over time for the launch rate given in Section \ref{case1}.}
    \label{ituactive}
\end{figure}

\begin{figure} 
    \centering
    \includegraphics[width=0.5\textwidth]{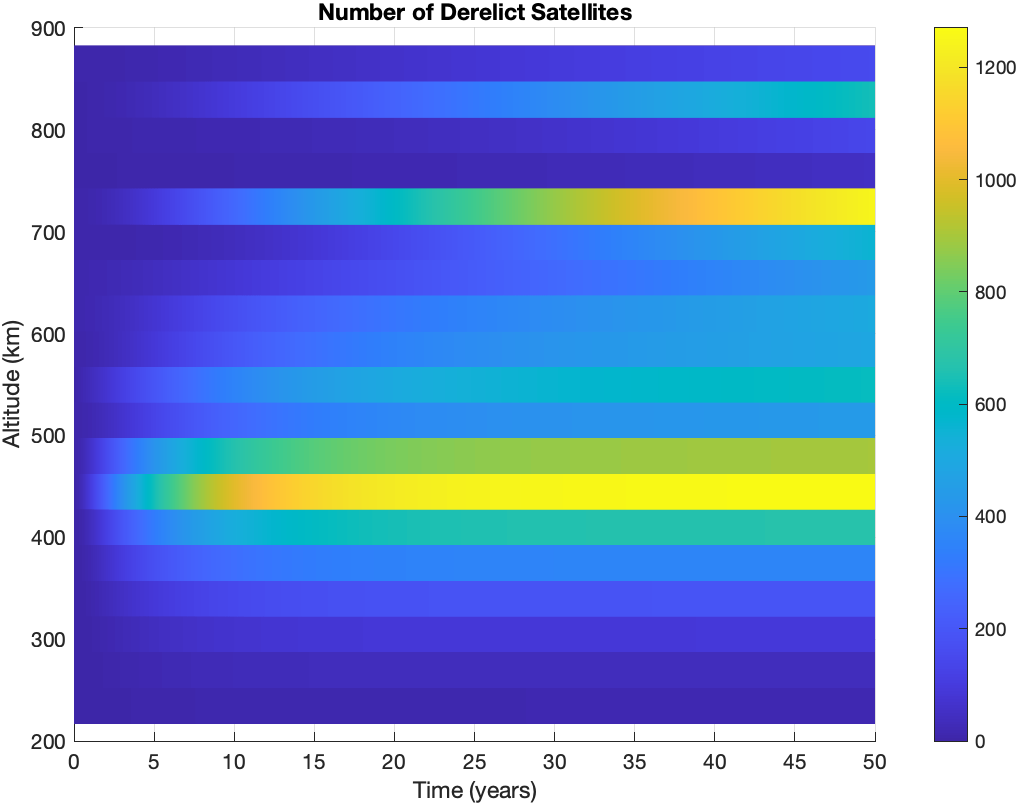}
    \caption{Number of derelict satellites in orbit over time for the launch rate given in Section \ref{case1}.}
    \label{ituder}
\end{figure}

\begin{figure} 
    \centering
    \includegraphics[width=0.5\textwidth]{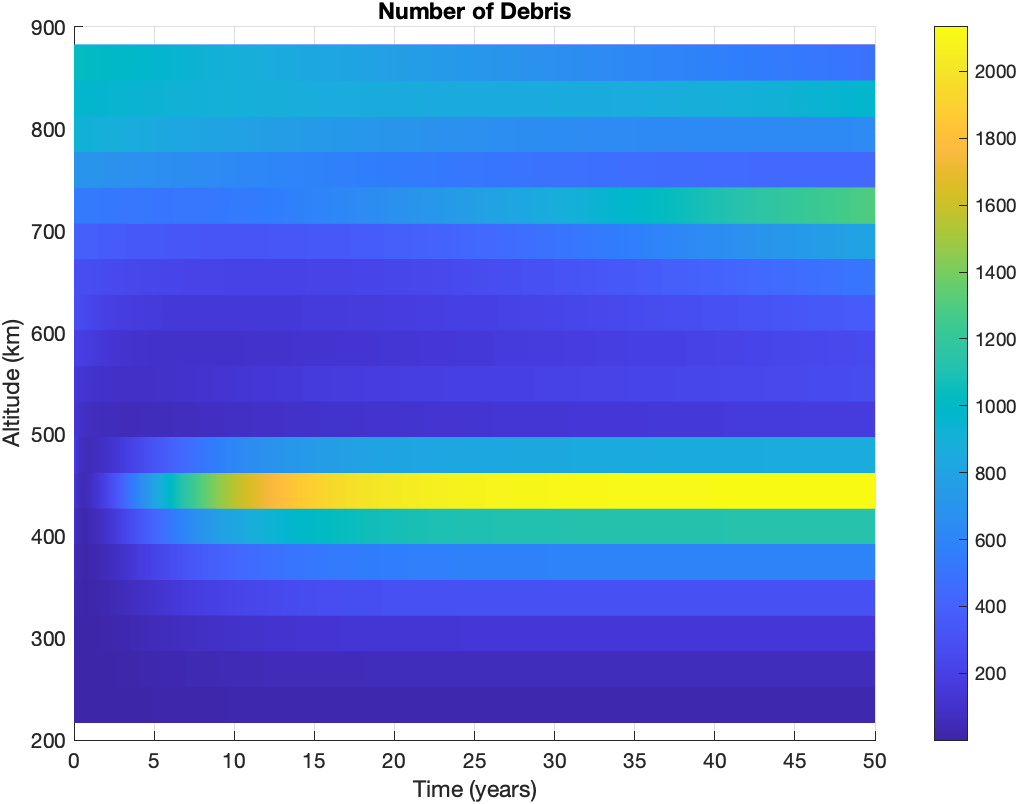}
    \caption{Amount of debris in orbit over time for the launch rate given in Section \ref{case1}.}
    \label{itudeb}
\end{figure}

We analyzed how many years were needed for the orbital environment to settle into its equilibrium state for the launch rate $\lambda_{itu} /21$ assuming the initial conditions are given by Figure \ref{initialfig}. We integrated the set of differential equations using these initial conditions. The results are shown for active satellites, derelict satellites, and debris in Figures \ref{ituactive}, \ref{ituder}, \ref{itudeb}, respectively. The population of active satellites settles into equilibrium across all shells within 10 years. The populations of derelict satellites and debris requires greater than 50 years to reach equilibrium for particular altitude shells. 



\subsection{Perturbing Launch Rate for One Year}
\begin{figure} 
    \centering
    \includegraphics[width=0.5\textwidth]{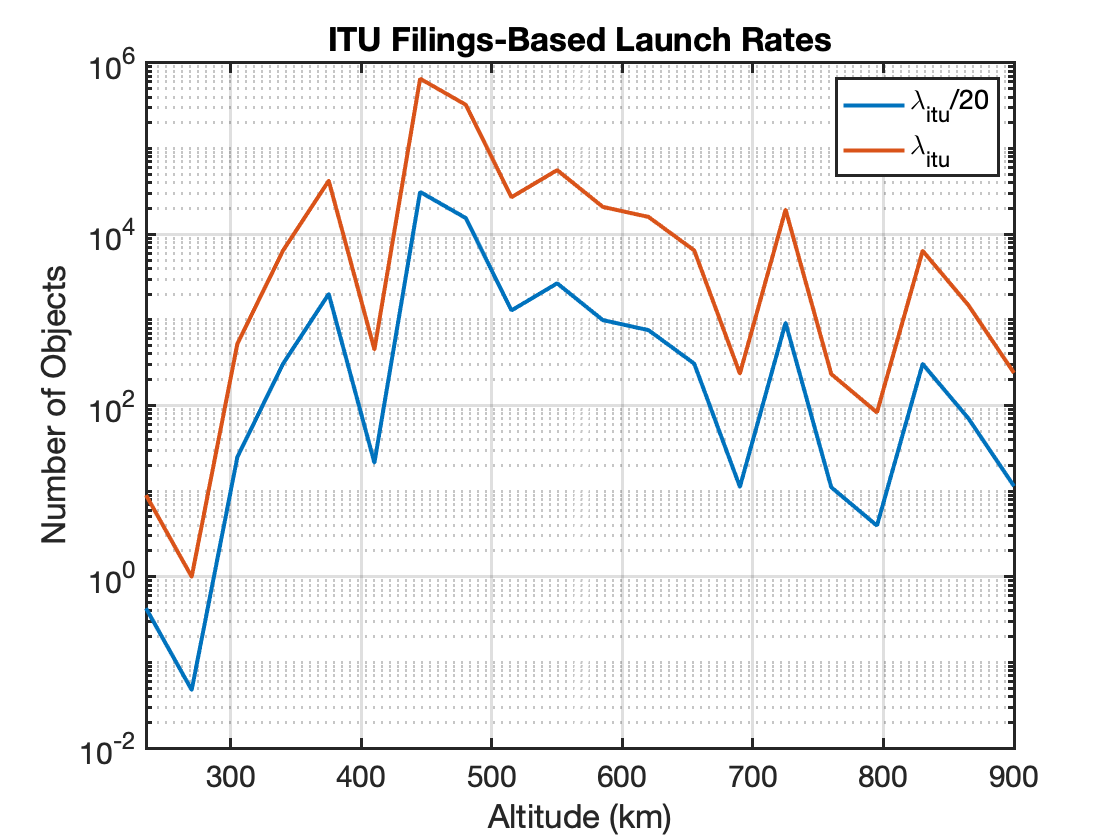}
    \caption{Launch forecast proportional to ITU 'As received' filings.}
    \label{twolaunches}
\end{figure}

We studied how the orbital environment would react to a one-time drastic increase in launch. 
This perturbation in launch rate allows for all ITU filings displayed in \ref{itufig} to be launched in one year. 
The utility of this approach is we can study how the orbital environment reacts to a launch rate for which no equilibrium solutions exist as stated in section \ref{equilibriumcalc}.  
We started with initial conditions given by Figure \ref{initialfig} and a launch rate of $\lambda_{itu} /21$ as shown in Figure \ref{ituequil}. We allowed the system of equations to evolve for 20 years at this constant launch rate and then we increased the launch by a factor of twenty for one year. These two launch rates are displayed in Figure \ref{twolaunches}.

\begin{figure} 
    \centering
    \includegraphics[width=0.5\textwidth]{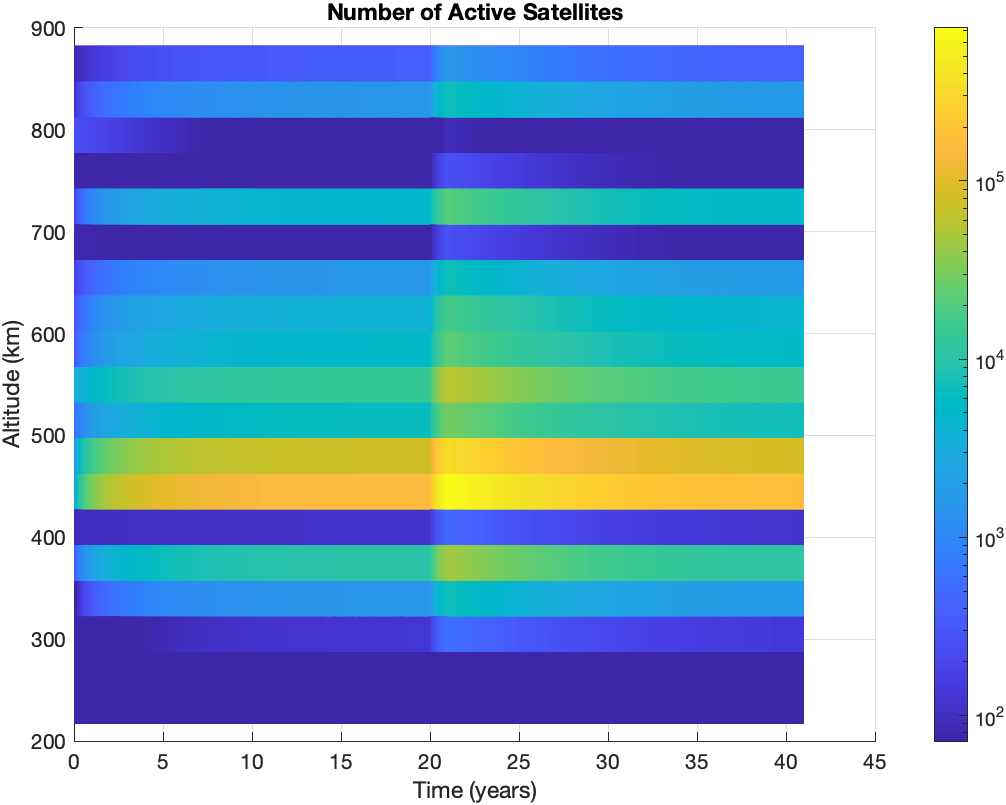}
    \caption{Number of active satellites over time with a one time increase in launch at 20 years.}
    \label{itupact}
\end{figure}

\begin{figure} 
    \centering
    \includegraphics[width=0.5\textwidth]{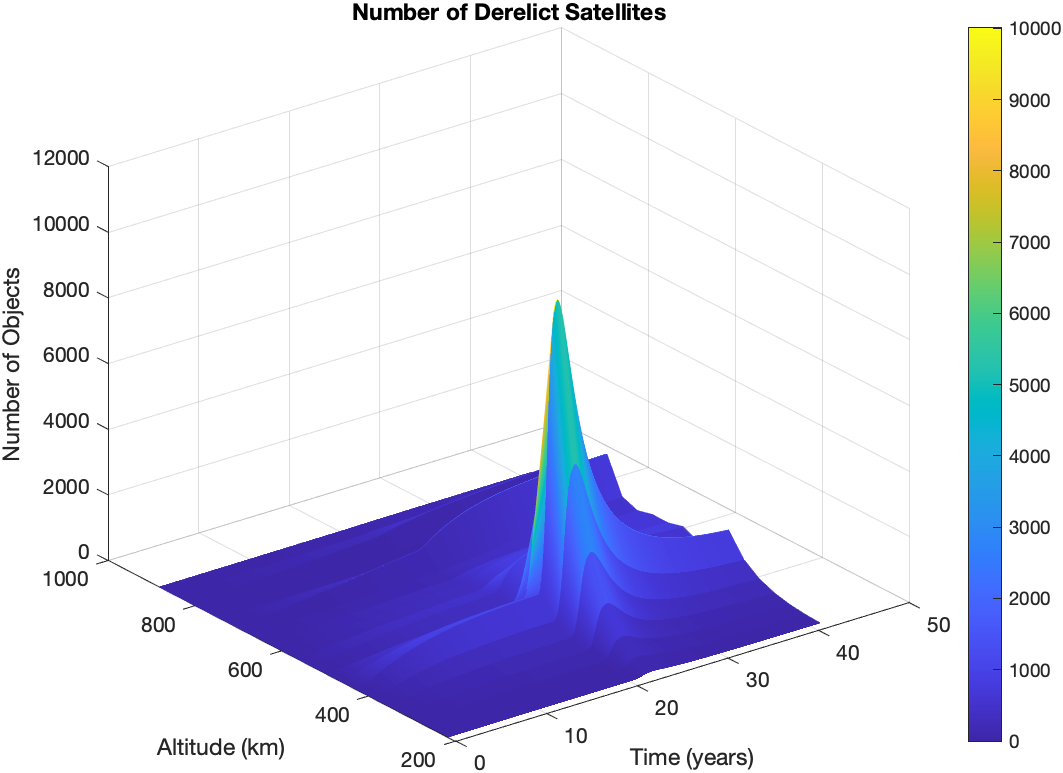}
    \caption{Number of derelict satellites over time with a one time increase in launch at 20 years. }
    \label{itupder}
\end{figure}

\begin{figure} 
    \centering
    \includegraphics[width=0.5\textwidth]{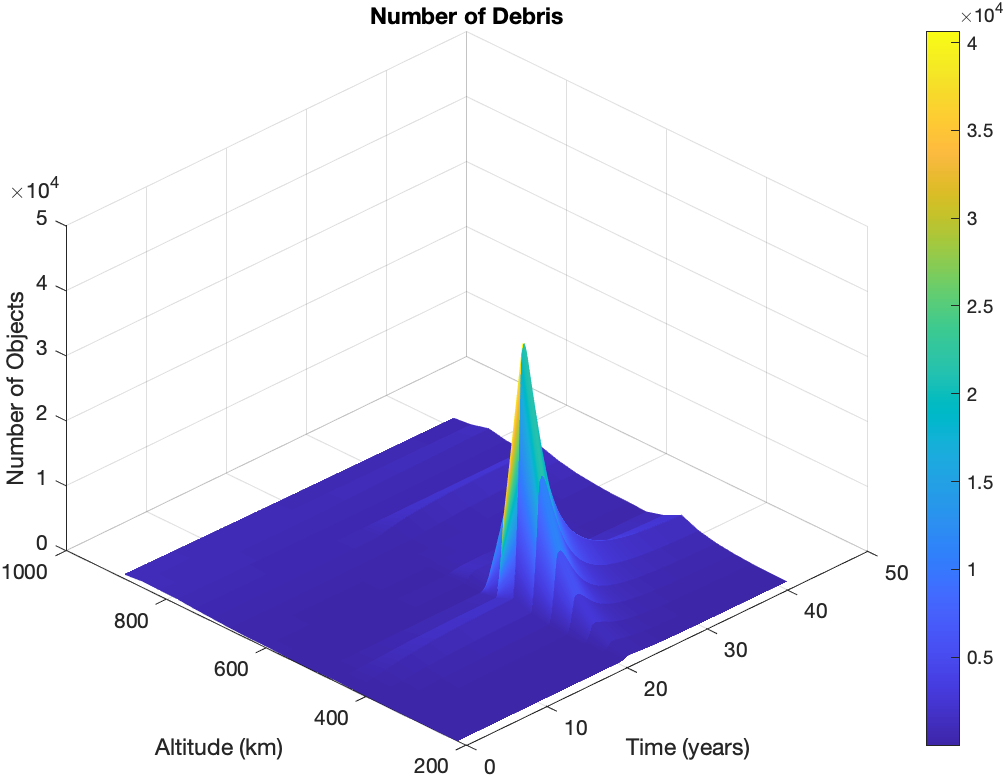}
    \caption{Number of debris objects over time with a one time increase in launch at 20 years.}
    \label{itupdeb}
\end{figure}

After this one year increase in launch activity, we decreased the launch rate back to the original rate of $\lambda_{itu} /21$ and allowed the system to evolve for another 20 years. 
The results are shown for each species in Figures \ref{itupact}, \ref{itupder}, \ref{itupdeb}. 
Overall, the system evolved back toward the stable equilibrium solution given in Figure \ref{ituequil} within the 20 years following the perturbation in launch rate. 
A change in the launch rate in effect perturbs the population of each species away from equilibrium. 


\subsection{Perturbations of Equilibrium Solutions}
We studied how the orbital environment reacts to the event of a sudden increase in debris. 
We analyzed two cases: the first case depicts the effect of an uniform increase in debris across all shells, and the second case depicts the effect of debris increase in one shell and compares a perturbation in debris at a high altitude vs. at a low altitude. 
\subsubsection{Equal Perturbation Across all Shells}
\label{equalshells}
\begin{figure}
    \centering
    \includegraphics[width=0.5\textwidth]{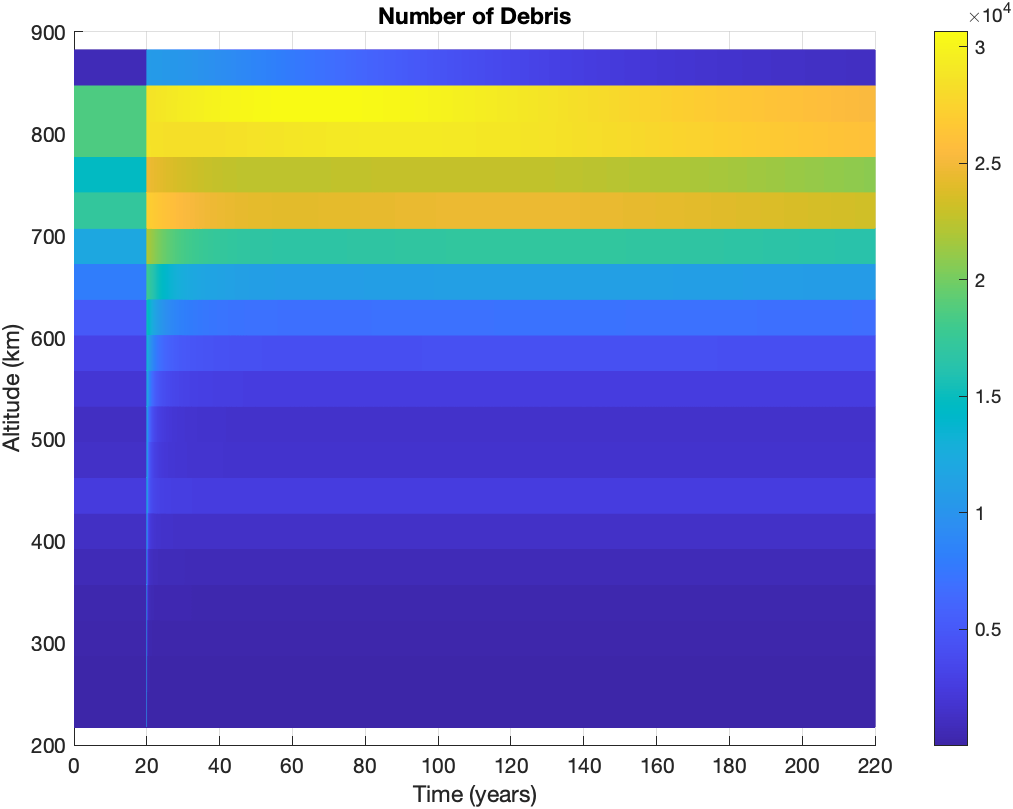}
    \caption{Debris population over time with an impulsive increase in debris by 10,000 fragments at 20 years.}
    \label{10000deb}
\end{figure}

Starting with the system at equilibrium depicted in Figure \ref{ituequil} for the constant launch rate given by case \ref{case2}, we perturbed the amount of debris in each shell by $10,000$ objects at $t=20$ years and allowed the system to evolve for 200 years. The change in the amount of debris is shown in Figure \ref{10000deb}. 
Table \ref{table10000} summarizes these results by displaying the total amount of each species before the increase in debris, at the time of the event, and 200 years after. 
From Figure \ref{10000deb}, we note that such an event creates a minimal effect in lower altitude shells with each species returning back to its equilibrium state within a few years. However, for higher altitude shells the scenario is drastically different with the system remaining out of equilibrium for at least $200$ years. 
This analysis shows how an `explosion' type of event that produces a large amount of debris across all shells, greatly affects the amount of debris present at higher altitudes for many years following the event. 

\begin{table}
\centering
\begin{tabular}{|p{1cm}||p{1cm}|p{1cm}|p{1cm}|p{1cm}|} \toprule
Species & Initial & At Event & After 200 Years & $\Delta$ \\ \midrule
S & 292830 & - & 292629 & -201 \\
D & 21290 & - & 22597 & +1308 \\ 
N & 107712 & 307712 & 145379 & +37667 \\ \bottomrule
\end{tabular}
\caption{Population of each species before and after a sudden increase in debris across all shells.}
\label{table10000}
\end{table}

\subsubsection{Impulsive Debris Perturbations in High Shell vs Low Shell}
\label{sepshells}
\begin{figure*}
    \centering
    \includegraphics[width=\textwidth]{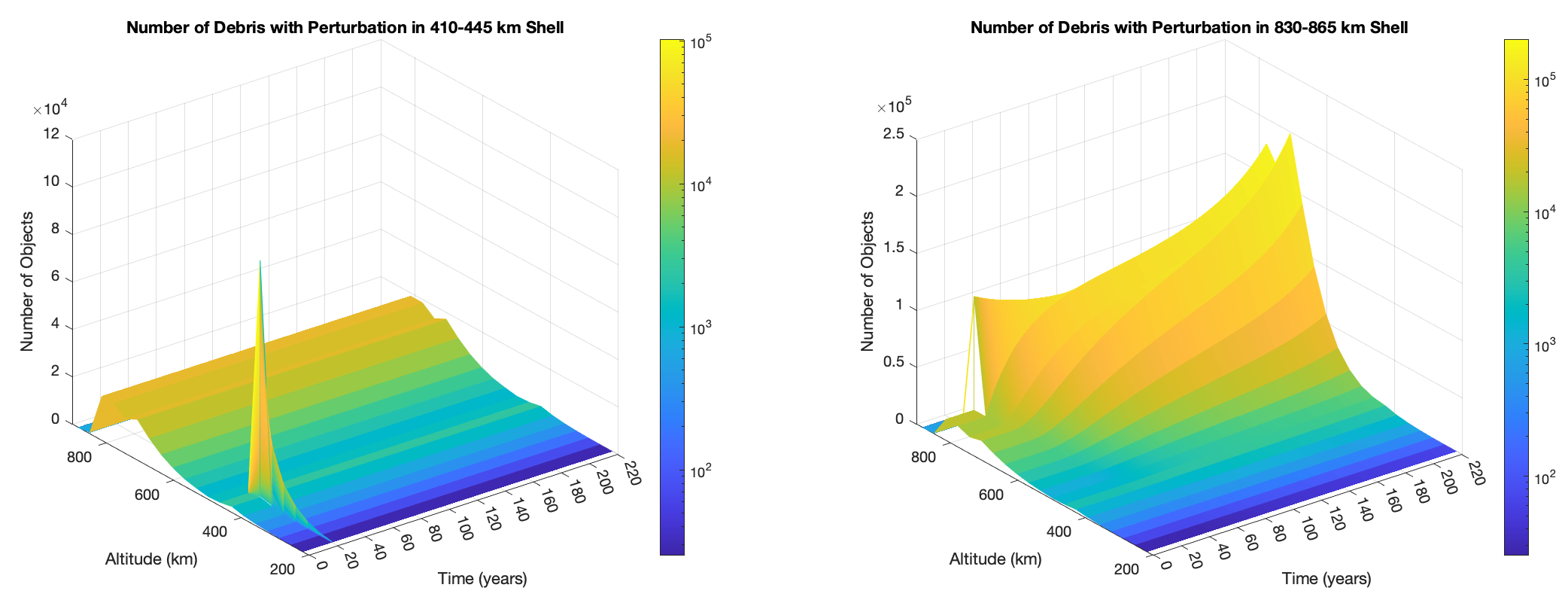}
    \caption{A comparison of the evolution of the debris population after a sudden increase in debris at $t=20$ years in two different shells.}
    \label{highlow}
\end{figure*}

From the equilibrium values displayed in Figure \ref{ituequil}, the amount of debris was increased by 10,000 objects in the second-highest shell (830km-865km) and in the shell with altitude range (410km-445km). We chose the second highest altitude shell rather than the highest shell since it contained significantly more active satellites. 
For the lower altitude shell, we chose the shell that contained the greatest amount of active satellites overall since this shell is the most sensitive to collisions between debris and active satellites.  
We set the orbital environment to equilibrium and then we added a perturbation in debris at $t=20$ years. After this perturbation, we allowed the system to evolve for 200 years. The results are displayed in Figure \ref{highlow}. Increasing the amount of debris by 10,000 objects in a high altitude shell has a much more significant impact on the overall LEO environment than increasing the debris in a lower altitude shell. 
In the left side of Figure \ref{highlow} we see that the environment quickly returns to its near-equilibrium state after the perturbation occurs, whereas the environment does not recover to equilibrium if such a perturbation occurs at a high altitude shell as shown in the right side of the figure. Rather than returning to its equilibrium state, a large perturbation of the debris population at a high altitude causes the population of debris to keep growing across multiple shells over the course of 200 years. It could be that for a longer period of time $t>200$ years, the system will return to its equilibrium state or it could be the case that the population of debris has reached an amount that causes collisions to continuously occur and debris to grow without end. We studied this behaviour in more detail in the next section \ref{kessler}.

\section{Instability Threshold: Kessler Syndrome}
\label{kessler}
Using the launch rate given in case \ref{case2}, we calculated the instability threshold as the maximum perturbation in debris away from equilibrium for which the population of debris continues to increase without bound over 1,000 years. The two types of perturbations we used were perturbing debris in all shells simultaneously as done in section \ref{equalshells} and perturbing debris in each shell individually similar to the approach used in section \ref{sepshells}.
The system does not necessarily need to return to its equilibrium solution within 1,000 years of the perturbation but for the perturbation to be considered apart of the stable region, the amount of debris and derelict satellites need to be decreasing at the end of the 1,000 years. In other words, at $t=1000$ years the set of $\{S,D,N\}$ must satisfy:
$$
\hspace{4pt} \dot{D}\leq0, \hspace{4pt}\dot{N}\leq0
$$
for all shells. The number of active satellites can be decreasing or increasing at the end time. In this way, we calculated the threshold at which run-away debris growth occurs, referred to as Kessler syndrome. 
\subsection{Perturbing All Shells Simultaneously}
\begin{figure*}
    \centering
    \includegraphics[width=\textwidth]{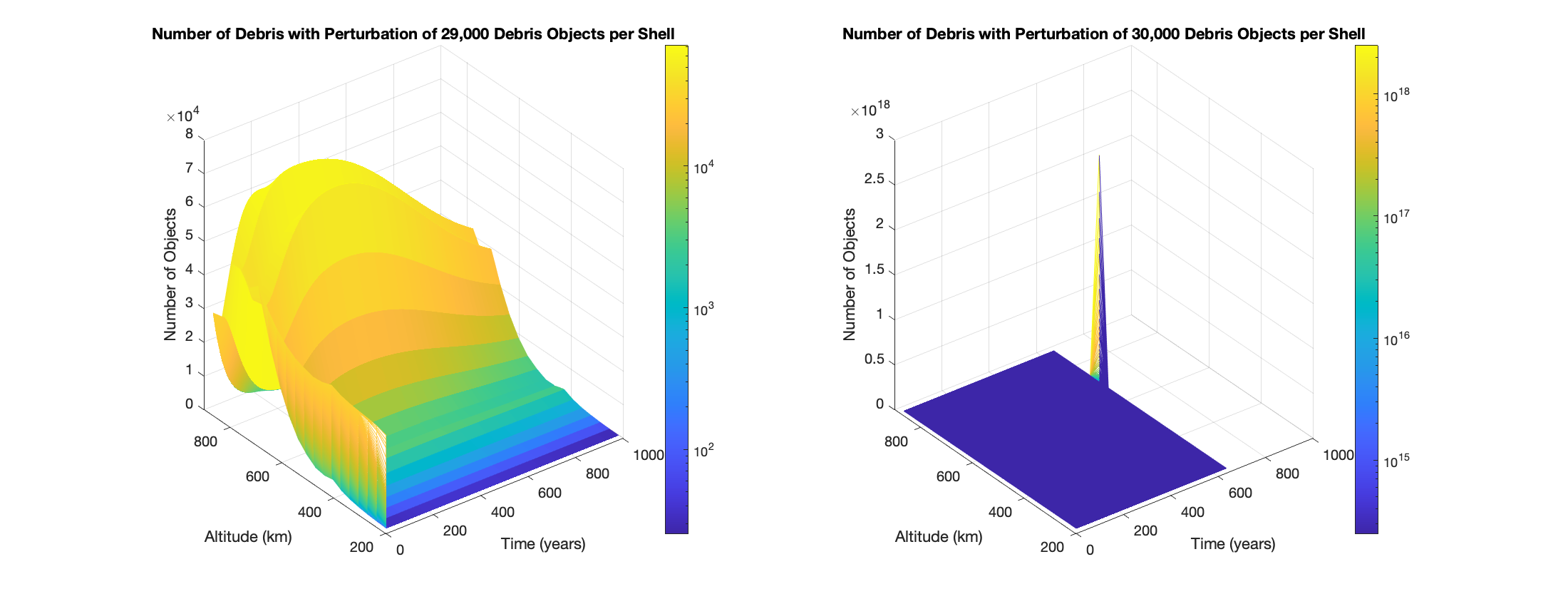}
    \caption{A comparison of the evolution of the debris population after a perturbation in debris occurs across all shells. The stable regime is displayed on the left and the unstable regime is displayed on the right.}
    \label{maxpet_allshells}
\end{figure*}

\begin{figure*}
    \centering
    \includegraphics[width=\textwidth]{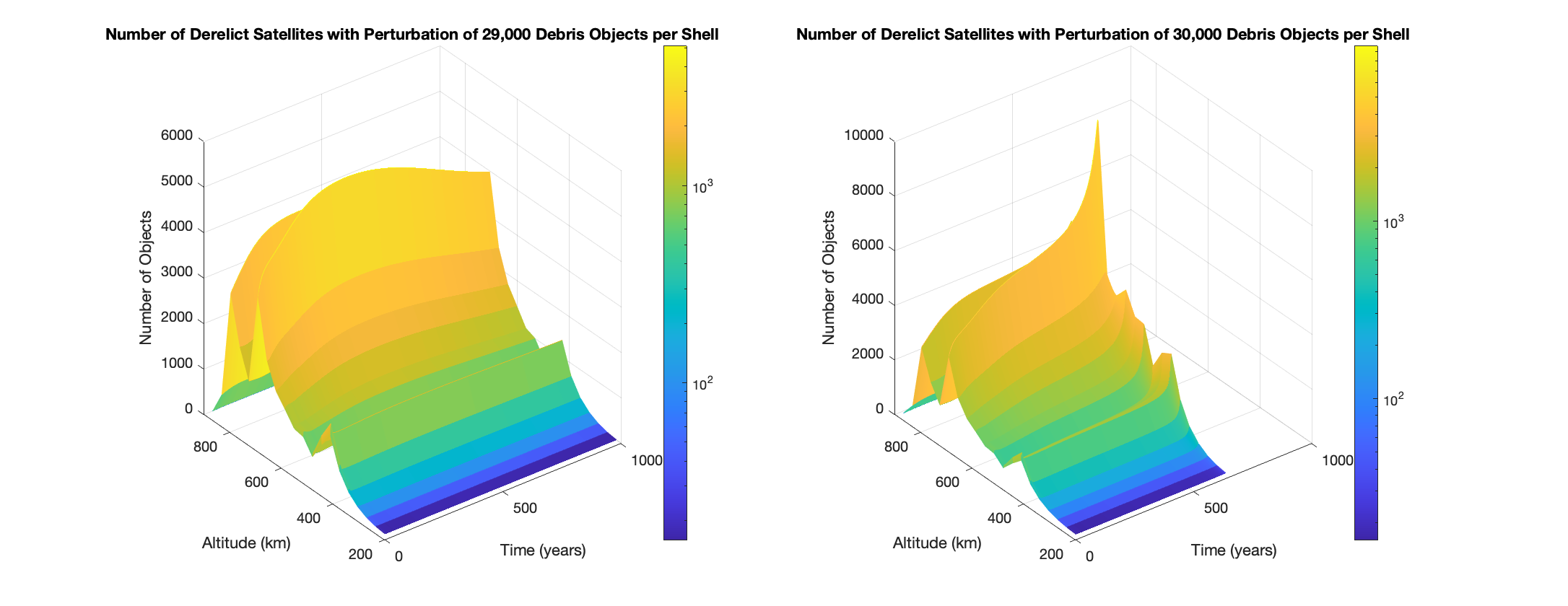}
    \caption{A comparison of evolution of the derelict population after a perturbation in debris across all shells.}
    \label{maxpet_allshells2}
\end{figure*}

To the nearest thousandth, the largest perturbation to the debris population for which the system reverted toward the equilibrium state after 1,000 years was found to be 29,000 debris objects.
Perturbing the debris population by 30,000 objects was found to cause runaway debris growth. These two cases are displayed in Figure \ref{maxpet_allshells}.
In the left of Figure \ref{maxpet_allshells} it is clear that the perturbation in debris causes debris growth for about 400 years but the system begins to return to equilibrium as it evolves for the remaining 600 years. 
This is not true for a perturbation of 30,000 debris objects as displayed in the right side of the Figure \ref{maxpet_allshells}, wherein the population of debris continues to grow reaching a couple quintillion before the integration fails at $t=600$ years. 
Such run-away debris growth displays Kessler syndrome as the orbital environment is unstable and continues to diverge away from its equilibrium state. Through collisions with active satellites, a perturbation in debris also causes a change in the population of derelict satellites. The evolution of the derelict population for a perturbation of 29,000 and 30,000 debris objects is displayed in Figure \ref{maxpet_allshells2}.
Overall, the runaway debris growth occurred in higher altitude shells, but we are also interested in the instability threshold of lower altitude shells. Thus, rather than simultaneously perturbing all shells away from equilibrium, we studied the instability threshold of each altitude shell in the next section.

\subsection{Perturbing Shells Individually}
By perturbing the amount of debris in each shell individually, we were able to analyze how sensitive each altitude shell is to a sudden increase in debris. We calculated the maximum perturbation in the debris population away from equilibrium to the nearest thousandth for which the system evolved back toward equilibrium within 1000 years. The instability threshold hence exists at this maximum perturbation amount.
The results are presented in Table \ref{tableshellpet}. Altitudes below 410 km are not included in the table because these shells could withstand a perturbation of debris equal to $10^8$ objects. We conclude that the stability of the orbital environment at lower altitude shells is much more resilient to perturbations in debris than higher altitude shells. 
This result concurs with the result of section \ref{sepshells}. 
The reasoning for this behaviour is the sink of the model, namely atmospheric drag, is much greater at lower altitude shells, which removes debris from the environment preventing collisions with active and derelict satellites that would create more debris. 
We would like to note that this analysis was done for a particular launch rate taken as a fraction of the ITU filings $\lambda_{itu}/21$ as shown in Figure \ref{ituequil}. 
Debris creation is directly affected by the launch rate since launch activity is the source of active satellites per shell, and a higher density of active satellites per shell creates a greater likelihood of collision with debris. Thus, different evolution of the debris population would occur for a different launch rate.  

\begin{table}
\centering
\begin{tabular}{|p{2cm}||p{1.5cm}|p{2.5cm}|} \toprule
Altitude Shell (km)&Debris at Equilibrium&Max. Perturbation in Debris \\ \midrule
410-445 &2462&53442000\\
445-480 & 1445&28478000 \\
480-515 &1145&17331000\\
515-550 & 1953&9861000\\
550-585 & 3077&5775000 \\
585-620 & 5085&3348000\\
620-655 &7983&2001000\\
655-690 & 12046&1216000 \\
690-725 &17185& 610000\\
725-760 &14725& 414000\\
760-795 &18644& 263000 \\
795-830 &18621& 91000\\
830-865 &703& 71000\\
865-900 &25& 63000 \\ \bottomrule
\end{tabular}
\caption{Maximum perturbation in debris per shell before Kessler Syndrome occurs.}
\label{tableshellpet}
\end{table}



\section{Remarks and Conclusions}
\label{sec:conclusions}
Given a launch rate distribution that is based on historic launch activities (\ref{case1}), the current orbital environment will evolve to a stable equilibrium state. 
In such a state, the sources of the model, namely due to launch and collisions, balance the sinks of the model, namely post-mission disposal and atmospheric drag. 
Thus if launch activities remain at current levels, Kessler syndrome will not occur in the $200-900$ km altitude range of the orbital environment, given the assumptions of our model. This is also true for an increased but constant launch rate (\ref{case2}) for which a stable equilibrium state exists. 
We note that the evolution of the environment from the current populations of active, derelict, and debris objects to this equilibrium state would take decades. 
Given a dynamic launch rate distribution, an ever-increasing launch rate entails the system will not reach an equilibrium state. However, if the launch rate becomes constant after a period of continuous growth then the system may evolve towards the equilibrium state if the growth rate was small enough. In our analysis, only a growth rate of 1\% in launch rate per year over 50 years lead to a stable equilibrium state. Larger growth rates in launch rate entailed no equilibrium state would be reached with each species population ever-increasing. 

Perturbations in the debris population away from equilibrium showed how sensitive the environment is to an increase in debris. 
In general, perturbations in the debris populations in higher altitude shells had more drastic consequences than perturbations in lower altitude shells due to increased atmospheric drag forces existing at lower altitudes. 
Run-away debris growth is more common at high altitudes, with Kessler syndrome resulting from  significantly smaller perturbations in debris than at lower altitudes. Thus a debris-generating event occurring at a high altitude is more dangerous than one occurring at a low altitude, since at higher altitudes such an event can trigger Kessler syndrome to occur.

\section*{Acknowledgments}
The authors would like to thank Thomas Roberts for sharing processed ITU filing data. The authors wish to acknowledge the support of this work by the Defense Advanced Research Projects Agency (Grant N66001-20-1-4028). The content of the information does not necessarily reflect the position or the policy of the Government. No official endorsement should be inferred. Distribution statement A: Approved for public release; distribution is unlimited. 

\bibliographystyle{ieeetr}
\bibliography{refs}

\end{document}